\documentclass[preprint2]{aastex}
\usepackage{times}
\usepackage{amssymb}
\usepackage{gensymb}

\usepackage{times}
\usepackage{txfonts}
\usepackage{bm}
\usepackage{color}
\usepackage{float}

\shorttitle{Coronae and Winds from Irradiated Disks in X-ray binaries}
\shortauthors{Higginbottom et al.}

\begin{document}

\title{Coronae and Winds from Irradiated Disks in X-ray binaries}

\author{Nick~Higginbottom}
\affil{Department of Physics \& Astronomy, University of Nevada, Las Vegas, 4505 S. Maryland Pkwy, Las Vegas,
NV 89154-4002, USA}
\email{nick\_higginbottom@fastmail.fm}

\author{Daniel~Proga}
\affil{Department of Physics \& Astronomy, University of Nevada, Las Vegas, 4505 S. Maryland Pkwy, Las Vegas,
NV 89154-4002, USA}
\email{dproga@physics.unlv.edu}



\newcommand{\ltappeq}{\raisebox{-0.6ex}{$\,\stackrel
{\raisebox{-.2ex}{$\textstyle <$}}{\sim}\,$}}
\newcommand{\gtappeq}{\raisebox{-0.6ex}{$\,\stackrel
{\raisebox{-.2ex}{$\textstyle >$}}{\sim}\,$}}


\label{firstpage}

\begin{abstract}
X-ray and UV line emission in X-ray binaries can be accounted for by a hot corona.
Such a corona forms through irradiation of the outer disk by radiation produced in the inner accretion flow.
The same irradiation can produce a strong outflow from the disk at sufficiently large
radii. Outflowing gas has been recently detected in several X-ray binaries via blue-shifted absorption
lines. However, the causal connection between winds produced by irradiation and the blue-shifted absorption
lines is problematic, particularly in the case of GRO J1655-40. Observations of this source imply wind densities about two
orders of magnitude
higher than theoretically predicted. This discrepancy does not mean that these `thermal disk-winds' cannot explain
 blue-shifted absorption in other systems, nor that they are unimportant as a sink of matter. Motivated by the inevitability of
thermal disk-winds and wealth of data taken with current observatories such as Chandra, XMM-Newton and Suzaku,
as well as the future AstroH mission, we decided to investigate the requirements to produce very dense winds.
Using physical arguments, hydrodynamical simulations and absorption line
calculations, we found that modification of the heating and cooling rates by a factor of a few results in an increase of the wind density
of up to an order of magnitude and the wind velocity by a factor of about two. Therefore, the mass loss rate
from the disk can be one, if not even two orders of magnitude higher than the accretion rate onto the central object.
Such a high mass loss rate is expected to destabilize the disk and perhaps
provides a mechanism for state change.
\end{abstract}

\keywords{accretion, accretion disks - hydrodynamics - methods: numerical - X-rays: binaries}

\section{Introduction}
\label{section:introduction}

X-ray binaries (XRBs) are systems in which a compact object such as a black hole (BH) or neutron
star (NS) accretes material from a secondary star. This material is believed to form an accretion disk surrounding the
compact object, producing intense X-ray radiation via both blackbody radiation from the inner regions of disk itself, and 
Compton up-scattering of lower energy photons from a hot corona. These systems are further subdivided 
into two classes, based upon their properties. High-mass X-ray binaries have O or B class secondaries, are 
fairly steady X-ray sources and are thought to have evolved in-situ
from a binary system. In contrast, low-mass X-ray binaries (LMXBs) have very faint secondaries, show 
dramatic changes in X-ray luminosity (X-ray bursts) and their evolutionary path is not clear. They may have evolved,
through mass transfer and loss, from a situation where the donor star was more massive \citep{2002ApJ...565.1107P},
or alternatively in dense regions like globular clusters it is possible for a lone NS or BH to capture a low mass 
companion \citep[e.g.][]{2006csxs.book..341V}.
LMXBs are highly variable systems, sometimes exhibiting an accretion disk-like spectrum (referred 
to as the `high/soft' state) and sometimes a power law spectrum (the `low/hard' state). Sources in the 
low/hard state usually also show a radio jet which disappears as the source transitions into the
high/soft state \citep[][and refs therein]{2004MNRAS.355.1105F}

Accretion disks are predicted to have a flared geometry, meaning that
the centrally generated X-rays will illuminate the surface of the disk, heating it and causing the surface
layers to evaporate
\citep[e.g.][]{shakura_sunyaev,1983ApJ...271...70B,1991ApJ...374..721K,1993ApJ...412..267R,1996ApJ...461..767W,
2002ApJ...581.1297J,2002ApJ...565..455P} forming a hot disk atmosphere.
The observational signature of this hot atmosphere is both X-ray 
and UV emission lines  \citep[e.g.][]{1984ApJ...278..270B,1993ApJ...412..267R,2005ApJ...625..931J} 
as the X-rays are reprocessed by 
not only the surface of the disk, but also the surface of the secondary \citep[e.g.][]{1990A&A...235..162V}.
This heating of the secondary has been proposed as a possible mechanism by which is mass transferred from
the star to the accretion disk \citep[e.g][]{1981ApJ...243..970L,1982ApJ...258..260L}.

As observations of XRBs have improved,
blue shifted absorption lines have been observed
 in more than a dozen high/soft state NS and BH LMXBs \citep[][and refs therein]{2013AcPol..53..659D}.
  These provide good evidence of the existence of outflowing material,
 most likely associated with the accretion disk, although the driving mechanism 
of these `disk winds' is a subject of ongoing discussion. 
A recent review of observations of XRBs is given by \cite{2013AcPol..53..659D} and we summarize their data here. 
They cite 
outflow velocities between 400 and 3000 $\rm{km~s^{-1}}$ in 30\% of NS systems, and 
 100 and 1300 $\rm{km~s^{-1}}$ in 85\% of BH systems. Photoionization modeling is usually employed to obtain
 estimates of the physical state of the absorbing gas, and such analysis gives a column density of between
 $\rm{4\times10^{22}}$ and $\rm{20\times10^{22}~cm^{-2}}$ and an ionization parameter of $2.5 \leq \log(\xi) \leq  4.5$ for NS
 LMXBs. The ionization parameter is a measure of the ionization state of the gas, and we use the common 
 definition 
 \begin{equation}
 \xi=\frac{L_x}{nr^2}
 \label{equation:xi}
 \end{equation}
 where ${L_X}$ is the ionizing luminosity, n is the number density of the gas, and r is the distance
 between the source of ionizing flux and the gas. 
 BH LMXBs have a wider range of properties, with a column density between
 $0.5\times10^{20}$ and $\rm{6\times10^{23}~cm^{-2}}$ and an ionization parameter of $1.8 \leq \log(\xi) \leq  6$.
 The ionization parameter is degenerate in density and distance, so in order to obtain information about the
size/distance of the absorbers, it is necessary to break the degeneracy by measuring the density of the absorbing
 gas. This has been done for the microquasar GRO J1655-40 (\citealt{2006Natur.441..953M,2008ApJ...680.1359M}
 but also see \citealt{2006ApJ...652L.117N}) and these measurements suggest a relatively high density 
 ($n\simeq10^{14}~\rm{cm^{-1}}$) which in turn implies a small radius.  Those systems
exhibiting absorption appear to be generally observed edge on \citep{2012MNRAS.422L..11P}, which implies an 
equatorial geometry
for the absorbing gas. As we will discuss later, this does not necessarily mean that any outflow is
also equatorial - it can equally well be a bipolar flow, that exhibits stratification in physical properties such as
density, ionization parameter or both.

Possible mechanisms to drive disk winds are magnetocentrifugal 
acceleration of gas guided by magnetic fields threading the disk 
\citep[e.g.][]{blandford_payne_82,1992ApJ...385..460E,2006Natur.441..953M}, radiation pressure 
acting on electrons or lines \citep[e.g.][]{1980AJ.....85..329I,1993ApJ...409..372S,1985ApJ...294...96S,
2002ApJ...565..455P} or thermal expansion of the hot disk atmosphere 
as a `thermal wind' when
the gas thermal velocity exceeds the local escape velocity \citep{1983ApJ...271...70B,1996ApJ...461..767W}. 
Whatever the mechanism, these winds are of great interest since they firstly provide a 
way in which the XRB can interact with its surroundings, and secondly, if the mass flow is large enough, they
could be the reason for the observed state change \citep[e.g.][]{2009Natur.458..481N,2012MNRAS.422L..11P,
{2013ApJ...762..103K}}. The fact that these absorption features appear in high/soft state sources but not in low/hard state
sources \citep{2013AcPol..53..659D} is further evidence that they are linked to state change.

In this work, we build upon earlier simulations of thermal disk winds \cite[][hereafter L10]{2010ApJ...719..515L}. 
L10 modeled the launching of a wind in a system based on  GRO J1655-40. 
In that work, a wind was launched but the velocity 
was too slow to account for the observed blue shifts of absorption lines and the density in the fastest parts
of the wind was lower than the observed values for GRO J1655-40. Although these results suggest that
thermal winds are unlikely to be the source of the
absorbing material in that system, thermal driving is still an important mechanism and deserving of further 
investigation. This is because it
is almost certain to operate at some level in LMXBs, 
and even if it is not the principal source of the X-ray absorbing gas, it may well be important in the overall evolution
of the system.
For example, L10 showed that significant mass would be lost from the disk by thermal winds, and this mass
loss is of the same order of magnitude
as that which would be expected to destabilize the disk and perhaps drive state change \citep{1986ApJ...306...90S}.
In addition, we cannot be sure if the wind in GRO J1655-40 is typical or extreme, 
and new observations which are likely come from the AstroH satellite \citep[][and refs therein]
{2014arXiv1412.1164D,2014arXiv1412.1173M} 
may provide
more examples. We intend to investigate whether, with modifications to the heating and cooling rates, we can produce a 
thermally driven wind with physical properties more in line with current observations, and provide a framework
which may be of use in understanding future observations.

The simulations we present here are all computed in the same way as described in L10, with three modifications.
Firstly, we consider only the optically thin case, so the radiation flux at any point in the simulation can be 
simply computed assuming a $1/r^2$ drop off from a centrally located source of X-rays. Secondly, we reduce
the computational domain size from $20R_{IC}$ in the original simulations to just $2R_{IC}$. 
$R_{IC}$ is the Compton radius, defined as the location in an accretion disk
where the local isothermal sound speed at the Compton temperature, $T_{C}$, of the illuminating spectral energy
distribution (SED) exceeds the local
escape velocity. The Compton radius is therefore given by
\begin{equation}
R_{IC}=\frac{GM_{BH}\mu m_H}{k_BT_C}
\end{equation}
where $M_{BH}$ is the mass of the central object, equal to 7$M_{\odot}$ for these simulations, $\mu$ is the 
mean molecular mass which we set to 0.6, and other symbols have the usual meaning. The Compton temperature
for the illuminating SED in these simulations is $T_C=1.4\times10^7~\rm{K}$ which gives a Compton radius of 
$\rm{4.8\times10^{11}~cm}$. In all these simulations, as in L10, the disk is assumed to be flat and thin - defined
via a density boundary condition at the midplane.

The change in domain size is motivated
by preliminary investigations which show that the acceleration zone for the wind was located inside $0.1R_{IC}$, in line
 with previous work \citep[e.g.][]{1983ApJ...271...70B,1996ApJ...461..767W,2002ApJ...565..455P}. In essence, this is because
 the most efficient acceleration of gas via thermal expansion occurs as the gas is heated past the lower equilibrium
 temperature on the thermal equilibrium curve (see Figure \ref{figure:stability_curves} and related discussion in the
 next section). This occurs at fairly low values of ${\xi}$, which 
 is where the gas is densest, i.e.  close to the central object.
The gas then enters an unstable heating zone where the 
 next stable temperature is over an order of magnitude 
 higher. Rapid heating occurs resulting in rapid acceleration as the gas expands. This change to the domain size
 means that we better simulate the densest parts of the wind where absorption is most likely to occur. We neglect the
 outer parts of the disk which means that our calculated mass loss rates are lower limits.

Finally, but central to this project, we will modify the heating and cooling
rates assumed in the heating term in the thermodynamic equations. We examine several cases, each with different
rates, and each
representing a modification to the thermal equilibrium curve. This changes the way in which 
the gas passes through the `heating' region, and we will see that this can have a profound impact on the velocity, 
density and hence mass loss rate of the wind. 

In the next section, we briefly discuss our methodology. We then discuss the details of the flows produced 
by the different heating/cooling parameters. Finally, we discuss the relevance of our results to the ongoing 
discussion of thermal wind in XRBs.

\section{Method}
\label{section:methods}

\begin{figure*}
\includegraphics{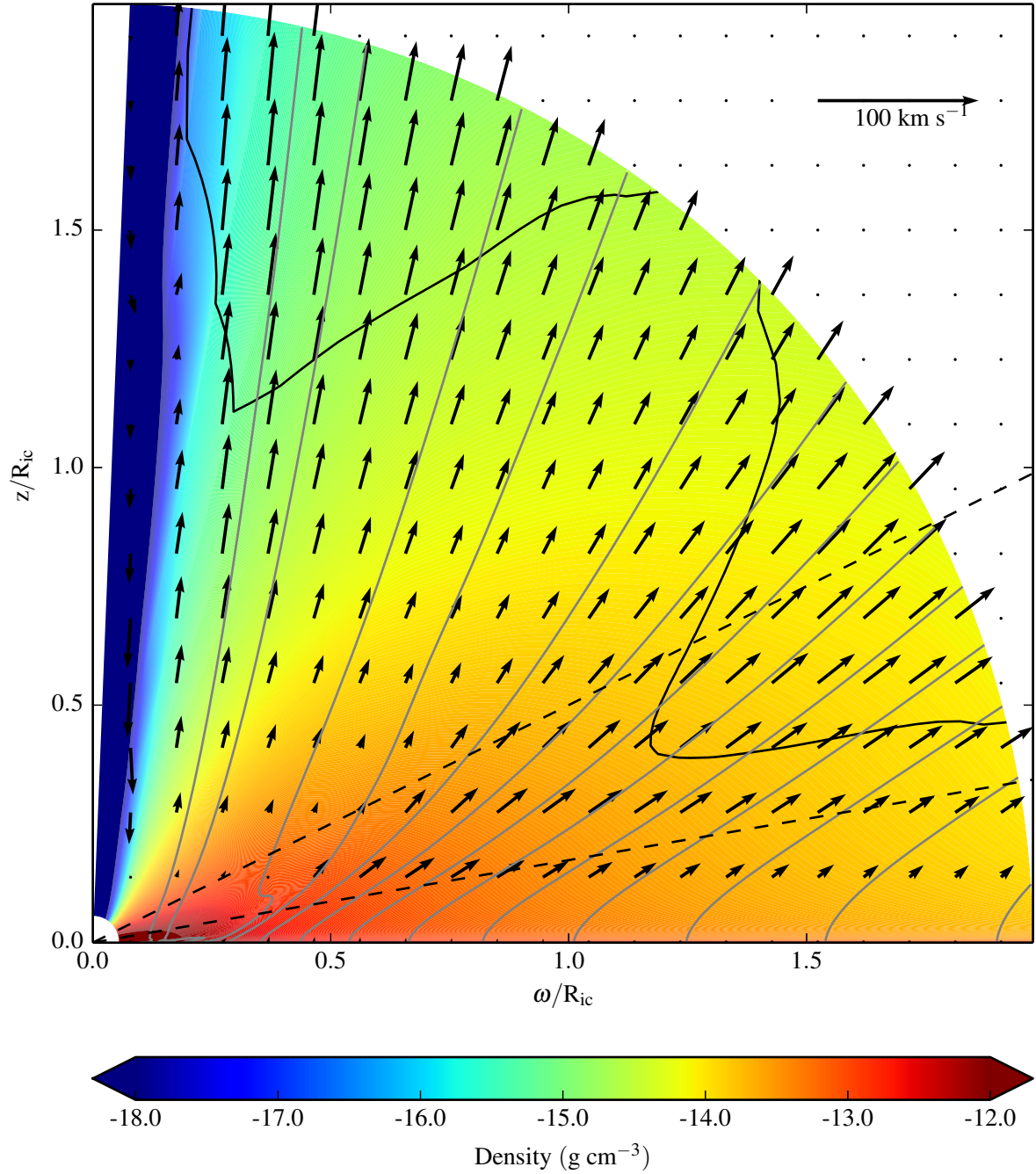}
\caption{The density  structure 
of model A (see Table \ref{table:wind_param}). Overplotted are streamlines (grey lines), the 80\degree~and 
60\degree~sightlines (dashed lines) and arrows showing the velocity field. 
Also shown is the location of the M=1 contour (black line). }
\label{figure:luketic}
\end{figure*}

Our simulations are based upon the simulations presented in L10 (run C8). Fig. \ref{figure:luketic} shows 
a rerun of that simulation, but computed upon a smaller grid, concentrating on the inner parts of the flow
where the acceleration occurs. The wind in this simulation, and all others presented here is shown
at a time of 220 000s which represents 25 sound crossing times (${R_{IC}/c_s}$), a time that is sufficient
for the wind to have settled
down to a steady state. Note that this timescale, about 2.5 days, also gives us an estimate of the lag between 
a change in the luminosity or SED of the central source of ionizing radiation and the response of the wind
through a change in structure. Of course changes in the ionization state of the wind would likely occur 
more quickly.

There are several important features of this model that can be seen in the
figure. Firstly, we see a dense, turbulent `corona' at ${\omega<0.3R_{IC},~z<0.2R_{IC}}$ where 
gas flow is severely inhibited by gravity and reaches steady state only in a time averaged sense. The streamlines
starting just outside this region \emph{can} escape, however it is still time dependent, hence the streamlines show `kinks'.

Secondly, outside the inner corona, the 
flow starts off moving vertically, as gas expands away 
from the accretion disk which lies along the ${\omega}$ axis. The streamlines bend outwards, partly due to 
the pressure gradient and partly as a result of conservation of angular momentum which means that the
centrifugal force acting on the rotating gas is not balanced by gravity. They are
self similar in this region. A fast, fairly dense flow
is produced which escapes at angles less than about 45\degree. Finally, we see a fast, low density infall
at polar angles. 

Although the L10 simulation did produce a fast fairly dense wind, as mentioned in the introduction
the density was 2-3 orders of magnitude
lower than that inferred from observations of GRO J1655-40. Even so, the total mass loss rate 
through the wind was 
seven times the accretion rate. 
We use the same version of the hydrodynamics code ZEUS-2D \cite{1992ApJS...80..753S} extended
by \cite{proga_stone_kallman} to carry out 2.5D simulations of the flow. In this code, radiative heating
 and cooling of the gas is computed using the parametrized rates first described by \cite{1994ApJ...435..756B}.
This parameterization includes Compton heating and cooling, photoionization heating, recombination 
cooling, bremsstrahlung cooling and collisional (line) cooling as functions of temperature T and ionization 
parameter ${\xi}$. The ionization parameter is defined as in Equation \ref{equation:xi}. The number
density of the gas is related to the density of the gas by ${n=\rho/\mu m_H}$ where the mean molecular 
weight ${\mu}$ is set to
0.6. Optical depth effects are not considered in the simulations here, so the factor of $L_X/r^2$ is related 
to the flux, reduced only by distance effects.

Compton heating and cooling is given by
\begin{equation}
G_{Compton}=8.9\times10^{-36}\xi(T_X-4T)~\rm{(ergs~s^{-1}~cm^{3})}
\end{equation}
where ${T_X}$ is the temperature of the illuminating power law spectrum (set to $5.6\times10^7~\rm{K}$).
 Photoionization heating and
recombinational cooling are subsumed into one term, referred to as ``X-ray heating/cooling'', given by
\begin{equation}
G_X=1.5\times10^{-21}\xi^{1/4}T^{-1/2}(1-T/T_X)~\rm{(ergs~s^{-1}~cm^{3})}
\label{equation:xray}
\end{equation}
whilst bremsstrahlung cooling is parametrized by 
\begin{equation}
L_b=3.3\times10^{-27}T^{1/2}~\rm{(ergs~s^{-1}~cm^{3})}.
\end{equation}
Finally, line cooling is given by 
\begin{equation}
L_l=\left[1.7\times10^{-18}\exp{\left(-T_L/T\right)}\xi^{-1}T^{-1/2}+10^{-24}\right]\delta
\end{equation}
where ${T_L}$ has the units of temperature and parametrizes the line cooling. It is set to $1.3\times10^5$ K. 
The ${\delta}$ parameter allows one to
reduce the effectiveness of line cooling due to opacity effects. In an optically thin plasma, ${\delta}$ is
set to 1. The units of ${L_l}$ are the same as for the other rates.

We are interested in investigating whether simple 
changes to the heating and cooling rates, thereby modifying the thermal equilibrium curve, can
increase the velocity and density of the wind to better
match observations. 
To modify these rates, we apply pre-factors to each of the mechanisms
and so the equation for the net cooling rate ${\mathcal{L}}$ $\rm{(ergs~s^{-1}~g^{-1}}$), 
which appears in the energy conservation equation, becomes
\begin{equation}
\rho\mathcal{L}=n^2(A_CG_{Compton}+A_XG_X-A_lL_l-A_bL_b).
\label{equation:heatcool}
\end{equation}

The first six lines of Table \ref{table:wind_param} gives the values of pre-factors, 
${T_X}$ and ${L_X}$ for the original
L10 simulation (run C8, denoted A) and six further simulations.
The line cooling pre-factor effectively replaces the ${\delta}$ parameter and therefore represents 
a measure of line opacity. Modifications of ${A_X}$, the photoionization/recombination rate pre-factor 
can be justified by a change to the illuminating  SED or metallicity of the gas. 
Calculations of 
the precise nature of the connection between these parameters and ${A_X}$ are beyond the scope of this
work, our values are not intended to represent a particular case, rather we adjust them to produce the 
desired thermal equilibrium curve. 
We change ${A_b}$ somewhat arbitrarily to make gas thermally stable everywhere. Our aim here
was to investigate the relationship between thermal instability (TI) and efficient acceleration. The upper and lower stable
temperatures remain the same and so we isolate the effect of instability with this experiment. 

The quantity $\xi_{cold,max}$ is the value of the ionization parameter when 
the flow becomes thermally unstable. \cite{1965ApJ...142..531F} demonstrated that in a non-dynamical flow,
subjected to isobaric perturbations, thermodynamic  instability results when
 $\left[\delta \mathcal{L}/\delta T\right]_p>0$. This is also where the gradient
${d\ln(T)/d\ln(\xi)}$ becomes greater than 1. Table \ref{table:wind_param} also gives $T_{eq}$, the equilibrium
temperature expected for $\xi_{cold,max}$. It can be shown that if line cooling is balanced by X-ray heating on the
cool branch of the stability curve, then this temperature is expected to be ${4/5~T_L}$. This is 104 000K,
and looking at the values in Table \ref{table:wind_param}, we see that the actual values
are very close to this. We also include values of $\Xi_{cold,max}$, which is the ratio of 
radiation pressure to gas pressure when the flow becomes thermally unstable. This is given by
$\Xi=F_{ion}/nk_bTc=\xi/4\pi k_bTc$ \citep{1981ApJ...249..422K}, where the temperature T is set to
the equilibrium temperature at the onset of instability.

To produce comparable simulations, we use a density boundary condition to 
ensure that the ionization parameter is equal to $\xi_{cold,max}$ at the Compton 
radius. The density at the midplane at the start of our simulations is defined by the equation
\begin{equation}
\rho(r)=\rho_0\left(\frac{r}{R_{IC}}\right)^{-2},
\end{equation}
where ${\rho_0}$ is given by
\begin{equation}
\rho_0=\frac{L_Xm_H\mu}{\xi_{cold,max}R_{IC}^2}.
\end{equation}
This is given in the table along with ${R_{IC}}$ for each case. Since the density in the disk is 
proportional to $1/r^2$, this means that ${\xi}$ is a constant in the disk plane, and it is the
same for all runs except Ah in which it is ten times bigger.

The hydrodynamic calculations are carried out in a spherical polar coordinate system,
running from 0 to 90\degree~ in angle, and from ${R_{in}}$ to ${R_{out}}$ in the
radial direction. The zone spacing increases with radius, such that ${ dr_{k+1}/dr_{k}=1.05}$
giving finer discretization in the inner parts of the flow. The zone spacing reduces with increasing
angle ${ d\theta_{k+1}/d\theta_{k}=0.95}$ giving more resolution close to the disk. These
parameters, together with the number of points used in the two dimensions are given in Table 
\ref{table:wind_param}.

\begin{table*}
\begin{tabular}{p{6cm}p{1cm}p{1cm}p{1cm}p{1cm}p{1cm}p{1cm}p{1cm}}
\hline Prefactors & A &Ah& B & C & D &E	 &F\\ 

\hline \hline 

$A_l$ & 1.0 &1.0& 0.2 & 1.0 & 1.0 & 1.0 & 0.076\\
$A_C$ & 1.0 &1.0& 1.0 & 1.0 & 1.0 & 1.0 & 1.0\\
$A_b$ & 1.0 &1.0& 1.0 & 3.9 & 1.0 & 1.0 & 1.0\\
$A_X$ & 1.0 &1.0& 4.0 & 1.0 & 1.0 & 1.0 & 1.0\\
\hline
\multicolumn{5}{l}{Physical Parameters}\\
\hline
$T_X (10^6~$K) & 56 & 56 & 56 &56 & 0.80 & 230 & 295 \\
$L_X (3.3\times10^{37}~\rm{ergs~s^{-1}})$& 1 & 10 & 1 & 1 & 1& 1 & 1\\
$log(\xi_{cold,max})$&  2.10 & 2.10 & 0.91 & N/A & N/A & 2.07 & 1.2\\
$log(\Xi_{cold,max})$&  1.33 & 1.33 & 0.17 & N/A & N/A & 1.32 & 0.43\\
$T_{eq}(\xi_{cold,max}) (10^3~\rm{K})$ & 111 & 111 & 106 & N/A & N/A & 109 & 113\\
$\rho_0 (10^{-12}~\rm{g~cm^{-3})}$ & 1.14 & 11.4 & 17.4 & 1.14 & 1.14 & 22.4 & 281 \\
$R_{IC} (10^{10}~$cm) & 48.2 & 48.2 & 48.2 & 48.2 &  3380 & 11.5 & 9.15 \\
\hline
\multicolumn{5}{l}{Grid parameters}\\
\hline
$R_{min} (10^{10}~$cm) & 2.4 & 2.4 & 2.4 & 2.4 & 2.4 & 1.2 & 1.2 \\
$R_{max} (10^{10}~$cm) & 96& 96& 96& 96& 96& 96& 96\\
$R_{ratio}$ &1.05&1.05&1.05&1.05&1.05&1.05&1.05\\
$N_R$ & 80& 80& 80& 80& 80& 80& 100\\
$\theta_{min}$ & 0.0& 0.0& 0.0& 0.0& 0.0& 0.0& 0.0\\ 
$\theta_{max}$ & 90.0& 90.0& 90.0& 90.0& 90.0& 90.0& 90.0\\
$\theta_{ratio}$ & 0.95& 0.95& 0.95& 0.95& 0.95& 0.95& 0.95\\
$N_{\theta}$ & 100& 100& 100& 100& 100& 100& 100\\
\hline
\multicolumn{5}{l}{Wind properties}\\
\hline 
$V_r(max~blueshifted)/100 ~\rm{km~s^{-1}}$ & 4.47 &   6.76 &   6.78 &   4.28 &  2.11 &  14.5 & 16.3\\
$V_r(\rho>1e12,max~blueshifted) /100 ~\rm{km~s^{-1}}$  & 1.18 &  1.98 &  3.26 &  0.203 &  0.186 & 5.03 & 2.51\\

$n_H~(60\degree~sightline)(\times10^{22}~\rm{cm^{-2})}$ & 5.71 & 53.0 & 74.2 & 1.01 & 0.281 & 13.3 & 15.0\\
$n_H~(80\degree~sightline)(\times10^{22}~\rm{cm^{-2})}$ & 46.3 & 476 & 441 & 22.6 & 24.1 & 16.6 & 255\\

$\dot{M}_{wind,disk} (\dot{M}_{acc})$ & 3.72 & 3.61 & 41.5 &  0.878 &  0.723 &  4.20 & 25.7\\
$\dot{M}_{wind,outer}(\dot{M}_{acc})$ & 3.45 &  3.36 &  38.3 &  0.638 &  0.758 &  3.85 & 24.9\\

\hline 
\end{tabular}
\caption{The heating and cooling parameters adopted in the simulations, and
some key parameters of the resulting winds.}
\label{table:wind_param}
\end{table*}
\begin{figure}[h]
\includegraphics{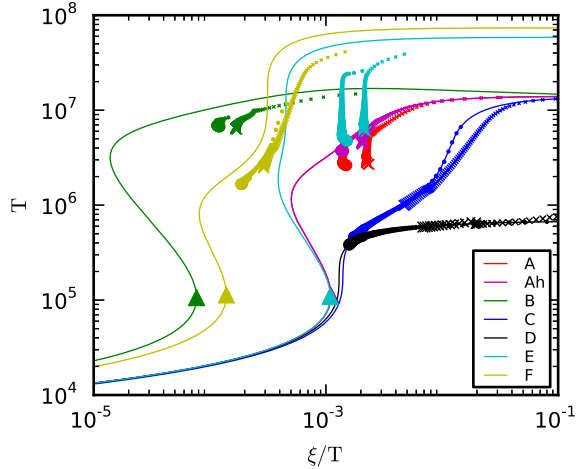}
\caption{Thermal equilibrium curves for the seven cases considered. 
The solid curves show the equilibrium temperature ($T$) vs ${\xi/T}$.
For each case, the crosses represent actual data from the 60\degree~ sightline and the
circles are from the 80\degree~ sightline. The size of the symbol shows
 the radial distance along the sightline, with larger symbols representing
the largest distances. The triangles show the point at which the heating
curve becomes unstable.}
\label{figure:stability_curves}
\end{figure}
The solid lines on Figure \ref{figure:stability_curves} show the thermal equilibrium temperature 
predicted for the
different cases plotted against ${\xi/T}$, which is equal to the ratio between the radiation 
and gas pressures. In an outflow, the gas pressure cannot increase along a streamline. This means that
${\xi/T}$ must always increase\footnote{Note that this is in fact only true in the case of weak or
zero magnetic fields. If magnetic fields are strong, then the gas pressure \emph{can} increase and the 
gas can move to the left on the equilibrium curve \citep{1992ApJ...385..460E,2000ApJ...537..134B}}, and so 
as a parcel of gas is heated, starting at the bottom left of the graph, 
it will follow these curves until
the gradient becomes negative (the onset of TI, if present). 
This location on the curve represents the maximum temperature of the cold
branch. The triangle symbol on the graphs shows this point. At this point, one expects the
gas to quickly heat up to the upper equilibrium temperature. This maximizes the rate of
energy transfer between the radiation field and the gas, driving expansion and hence 
acceleration. Thus it is in the unstable zone where the most `efficient' acceleration
of the gas takes place in order to form an
outflow.

The behavior of the different 
cases is best explained in the context of the shape of these thermal equilibrium curves.
Case A and Ah have identical thermal equilibrium curves, becoming unstable at the same value of 
${\xi}$ and reaching the same upper equilibrium temperature (set by the balance of
Compton heating and cooling). The difference in luminosity between the two cases has
no effect on the shape of the equilibrium curves, however as we will see in the next section, 
it does affect how the gas is heated. Given the same unstable zone, but enhanced radiation
density, one would expect case Ah to produce a faster, denser flow  \citep[see][]{1983ApJ...271...70B}.

Case B has reduced line cooling and enhanced photoionization heating. This means that
the gas becomes unstable at a lower ${\xi}$. Changes to the SED have been shown to
have this effect \citep[e.g., see][who use detailed photoionization calculations]{ 2000ApJ...537..134B}, 
so our approach is not unreasonable.
 The distance to the radiation source will be
largely unchanged, so low ${\xi}$ equates to higher density 
and so one would expect that the resulting outflow would be denser. 

Case C and D are both attempts to remove the unstable zone entirely. Case C has strongly 
enhanced bremsstrahlung cooling, through an increase in ${A_b}$. We did this, somewhat
arbitrarily, so that there is no TI for ${10^5\leq T\leq10^6}$ but the equilibrium temperatures
at the low and high end are the same as in A, Ab and B. Our goal here was to isolate the 
importance of TI to acceleration of the gas. Case D achieves the same thing,
 by reducing the X-ray temperature.
This reduces both the initial photoionization heating rate, and also the upper
equilibrium temperature.

Finally, cases E and F represent solutions with two unstable zones. In case E, we simply 
increase the X-ray temperature. This does indeed produce two unstable zones, however the
second unstable zone is at a lower pressure than the first. In case F, we increase the 
heating rate on the lower branch in a similar way to case B, by reducing line cooling. This
shifts the lower unstable zone to lower ${\xi}$ but leaves the upper unstable zone
unchanged. The aim for these two cases is to see if one can get a faster wind by extending the 
acceleration zone. It is also interesting to see if gas `collects' at the stable zone between the 
two unstable ones.

\section{Simulation results}
\label{section:sim_res}
We present the results of our simulations in the context of the thermal equilibrium curves. 
This allows us to see how the hydrodynamics of the winds affects the thermal balance, 
and thus give insight into how the winds are accelerating. The symbols, plotted over the solid
 lines, on Figure \ref{figure:stability_curves} show the relationship between the temperature
 and ionization parameter divided by that temperature
 in a range of cells along two sightlines. We can therefore see if the equilibrium temperature is
 reached in the simulations.

We note that there are no points on the lower stable branch of the 
stability curves for any of the cases. This is by design, since points along the lower branch
are essentially in the disk, and merely exhibit turbulent motions.
We have carried out detailed resolution studies using 1D simulations which 
demonstrated that the behavior of the wind in the 
regions sampled by our sightlines is not dependent on resolving the transition from cool stable
branch to unstable zone. Our limited 2D resolution study confirmed this.

Turning our attention to individual cases, we first examine cases A and Ah. The increase
in luminosity has had the desired effect, in that all of the points for case Ah lie on the upper
branch of the unstable zone, whilst those of case A lie below the curve. Adiabatic cooling as the 
hot gas expands holds the temperature below the 
expected stable temperature. This indicates that more energy has been transferred to the gas in the
higher luminosity case. Note however that the points from the two simulations occupy remarkably
 similar locations in ${T-\xi}$ space, 
given that Ah has an order of magnitude higher luminosity. This is because the density of
the wind has increased accordingly, giving a very similar value of the ionization parameter for
both flows.

Case B has, as expected, produced hotter gas at lower ${\xi}$. However, as
with case A, the gas is cooler than the upper stable temperature. This suggests that with a higher luminosity, 
one could obtain even hotter gas. Interestingly, the 80\degree ~points are all very close, indicating
that we are sampling points at the same relative distance along each streamline that the 
sightline intercepts. 

Cases C and D, where there is no unstable zone, the points tend to lie along the thermal
equilibrium curves 
with some exceptions. The innermost 60\degree sightline data of case C are cooler than expected if 
radiative processes dominated the heating/cooling. This is because adiabatic expansion is acting as 
an additional cooling mechanism. 
This simulation fails to launch a wind in the current simulation domain.
Case D also fails to launch a strong wind
and the polar infall which is always seen in these simulations extends to much larger inclination angles.
Compression of the gas by this flow at low radii produces some gas that is heated above the upper
stable temperature. This occurs for gas with an ionization parameter greater than about 10, and is
not shown on Figure \ref{figure:stability_curves}.

Case E shows two new effects. Firstly, despite there being two unstable zones,
the gas is unable to access the upper unstable zone since this lies at higher pressure. Therefore
the gas jumps straight from the lower unstable point towards the upper stable branch. Adiabatic cooling 
prevents the gas from reaching the upper branch and the existence of gas in a 
formally prohibited part of the thermal equilibrium graph demonstrates that hydrodynamic effects are 
an important consideration in the thermal balance of this type of wind. It is often assumed that
gas will avoid unstable zones \citep[e.g.][and refs therein]{2013MNRAS.436..560C}, 
providing an explanation of why some ionization states are not seen
in observations. This simulation demonstrates that the situation may be more complex.

Finally, case F does produce points close to the stable zone between the two instabilities since
that part of the curve is now physically accessible to the flow. 

In the lower section of Table \ref{table:wind_param} we provide some of the physical properties of the 
simulated outflows. First, we list the maximum radial velocity seen in each of the simulations. It
is of the same order of magnitude (a few hundreds of kilometers per second) as the blue shifted 
absorption lines seen in LMXBs \citep[e.g.][and references therein]{2013AcPol..53..659D,
2013AdSpR..52..732N}, however the velocity of gas is not
the only important factor. 

The density is also important since only dense gas will produce observable
absorption. The maximum velocity in regions with a particle density greater than
 $\rm{10^{12}~cm^{-1}}$ is much smaller. This is simply because we are probing regions deeper into the 
 wind, where the flow is still accelerating, and we see that the two stable cases do not produce fast
 enough gas at high density. Cases B, E and F do show fast moving gas with density above the
 threshold density. In cases E and F, the fast, dense gas is limited to a very narrow range of angles, 
 in case E it is within 3 degrees of the disk and would therefore probably not be observable. In  
 case E, the fastest dense gas is at small radii around ${\theta=75\degree}$. Whilst this could
 in principle be observed, the small angular range over which an absorption feature would appear
 would likely mean it would be transient.

 Another important physical parameter that can be derived from observations is the total column 
 density. This is between about $10^{20}-10^{23}~\rm{cm^{-2}}$ for all kinds of LMXBs 
 \citep{2013AcPol..53..659D}. Our simulations produce column densities of the right order of magnitude
 for equatorial sight lines, and indeed the 80\degree sightline would be Compton thick in case Ah and B.
 
 Finally, we give the mass loss rate, both leaving the disk ($\dot{M}_{wind,disk}$) and leaving
the computational domain ($\dot{M}_{wind,outer}$). These rates are calculated directly
from the simulation results, using $\dot{M}=\sum A\rho v$ where $A$ is the area represented by a cell, either
on the disk or at the outer boundary, $\rho$ is the density of the cell, and $v$ is either the vertical velocity 
for disk cells, or the radial velocity for cells at the outer boundary. The summation is carried out
over all relevant cells.
We report these values in terms of the accretion
rate, $\dot{M}_{acc}=4.4\times 10^{17}~\rm{g~s^{-1}}$ (assuming an efficiency of 8.3\%).
In L10, their version of model A produced an 
outflow of about 7 times the accretion rate whereas we see an outflow rate of about half that much. This
is simply because our simulation has a much reduced radial extent compared to the  L10 run. Increasing the luminosity
(model Ah) increases the mass loss rate, but the ratio of mass loss to accretion rate remains the 
same. Cases B and F produce significantly higher mass loss rates, in excess of the threshold of $15\dot{M}_{acc}$
that \cite{1986ApJ...306...90S} demonstrated could induce oscillations in an accretion disk. 
It should be noted that the winds are emerging from the disk
far outside the radius where most of the radiation is produced. Almost all of the ionizing radiation from a 
thin disk in a system like this one is produced within 100 gravitational radii of the centre. By comparison, 
the induced Compton radius is about half a million gravitational radii. Therefore, at least for the purposes
of these simulations, the model of a point-like, unvarying source of radiation at the centre of the simulation
is not necessarily invalidated by the prediction of large mass losses from the outer parts of the disk, at least over the
timescale of the simulation.

\section{Synthetic absorption line profiles}
\label{section:spectra}
It is useful to produce synthetic line profiles for our simulations, in order to get some idea of whether
the outflows could, in principle, produce the absorption features observed in XRBs. Computation
of the ionization state and level populations of the gas is beyond the scope of this work, and
we calculate the absorption using the simplified scheme discussed below. This scheme takes account
of thermal line broadening, and the doppler shifting due to the bulk flow of the wind, however we do not model
line emission here.

The opacity due to a resonance line (uncorrected for stimulated emission) is given by
\begin{equation}
\alpha(\nu)=\frac{h\nu}{4\pi}n_1B_{12}\phi(\nu)
\end{equation}
We arbitrarily set $B_{12}=1$ and use the hydrogen number
density for $n_1$. Therefore, our line opacity becomes
\begin{equation}
\alpha_{\nu}=\frac{h\nu}{4\pi}n_H\phi(\nu).
\end{equation}
For the line shape $\phi(\nu)$ we use a gaussian line profile of the form
\begin{equation}
\phi(\nu)=\left(\frac{c}{\nu_{0}}\right)\sqrt{\frac{m}{2\pi k_BT}}exp\left(-\frac{mc^2\left(\nu-\nu_0\right)^2}{2k_BT\nu_0^2}\right)
\end{equation}
For each radial cell i, we compute the line opacity as a function of frequency, and then doppler shift that line
profile to take account of the bulk flow velocity. We then
obtain the total, frequency dependent optical depth by summing up the opacity at each
frequency due to each cell of radial thickness dr,
\begin{equation}
\tau(\nu)=\sum_{i=inner}^{i=outer}\alpha_i(\nu)dr.
\end{equation}
This sum is computed from the innermost radial cell to the outermost, thus making the implicit assumption 
that the continuum source is point-like, and located at the origin.
A measure of the absorption profile of a generic line is then computed 
\begin{equation}
F(\nu)=e^{-\tau(\nu)}.
\end{equation}
Since no attempt is made to compute the density of any given ionic species,
these spectra are in no way accurate representations of what we would
expect to observe for a given system. Rather, they are just a means of comparing runs, 
since each spectrum is calculated in a consistent way.
\begin{figure*}
\includegraphics{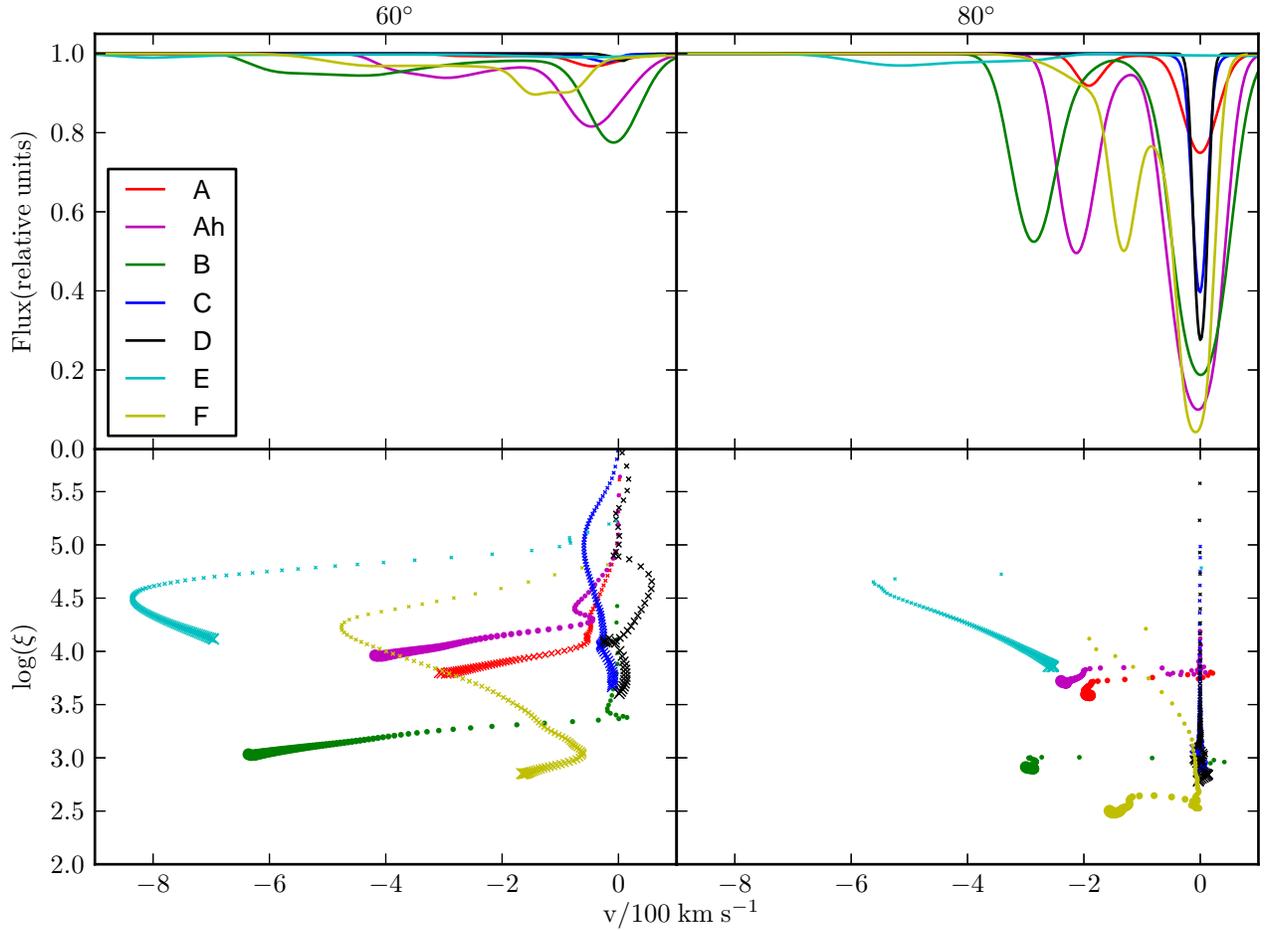}
\caption{Simulated spectra for the 60\degree (left hand upper panel) and 80\degree (right hand
upper panel) sight lines. Lower panels show ionization parameter vs radial velocity for all cells
in the two sight lines. The size of the symbol represents the distance from the central mass (
small symbols are for small distances). Crosses represent cells with a number density of less 
than $\rm{10^{12}~cm^{-3}}$
and filled circles show cells with a density above this threshold. Negative velocities represent 
motion away from the centre of the simulation, and thus blue shifted absorption. Note that
in the bottom right hand panel, the black and blue points overlie one another.}
\label{figure:simulated_spectra}
\end{figure*}

The upper panels of Figure \ref{figure:simulated_spectra} show the results of the line profile calculation
for the 60\degree~ (left panel) and 80\degree~ (right panel) sightlines. We immediately notice that the 80\degree~
features are much deeper than those seen at 60\degree. This is simply due to the higher density
at the base of the wind. This means that the ionization parameter is also generally lower in the
80\degree sightline, and so it is likely that different species would be observed at the two angles. 

A general feature of most of the spectra is an absorption feature close to zero velocity, and a second 
feature at a blueshift that varies from model to model. Although absorption features are seen
at zero velocity in observations, it is fair to say that the very strong features we predict here are
not commonly observed. We know however that by including line emission these absorption features would 
be weakened.
We are more interested in the second absorption feature which appears
at velocities between about $\rm{-150~km~s^{-1}}$ and $\rm{-800~km~s^{-1}}$.

Looking in detail at each of the cases now, we first see that the increased luminosity of case Ah
over case A was partially successful in that the absorption is stronger. This is of course due to
the increased density of the outflow. This can be seen clearly in the lower left panel, where the cells in case A
have density less than $\rm{10^{12}~cm^{-3}}$ (represented by the cross
symbol) whereas the same cells in case Ah have density greater than $\rm{10^{12}~cm^{-3}}$ (circles).
However, increase in density was not our only aim, we also
wanted to increase the velocity at which the absorption was seen. In this we have been only
partially successful, with the blue shifted absorption feature shifting to only very slightly higher 
velocities.

Case B is perhaps the most successful new model with blue shifted absorption features seen in 
both the 60\degree and 80\degree sightlines. The blue shifted absorption feature at 80\degree~ is 
deep and its width is only
due to thermal broadening. This is because all of the cells producing the feature are at nearly the same
velocity. This is in turn because photons flying along this sightline encounter gas in a very similar
physical regime at all radii showing that, close to the disk at least, the gas is flowing along highly 
self similar streamlines in this case. Since the ionization parameter is set to be the same at 
all radii at the mid plane, all gas will start in the same physical state. In case B,
the physical state of the gas has evolved similarly at all radii, `remembering' its initial conditions, up to the 80\degree
point. In contrast, by the time it has moved up to the
60\degree sightline, that `memory' of the starting state has been lost, and gas at different radii is in 
different physical conditions. Thus the absorption is produced by cells at a range of velocities and so the feature 
is much shallower 
but very broad.

As already discussed above, case C and D fail to produce fast outflows. This is clearly shown in the lower
panels, where all the cells from these simulations are clustered around zero velocity. These two
cases produce relatively narrow (the temperature of the gas is lower) features at zero velocity. 

As shown in Table \ref{table:wind_param} cases E and F do both produce dense, fast moving 
material. However, the material in case E is very close to the accretion disk, and is missed
by both sight lines shown here. Appearing at angles greater than 87\degree, it is unlikely
that it would be observable in any case. Case F does produce fast material at lower 
 ${\theta}$ and this is seen in the spectra as absorption around $\rm{-200~km~s^{-1}}$.
 Faster gas does exist in the simulation, but only over a very narrow range of angles.

\section{Discussion}
\label{section:discussion}

Our aim was to see if simple, physically motivated changes to 
the heating and cooling balance of the thermal wind simulated in L10 could produce a wind model that was 
more in line with observations. There are three main observational measures, the line velocities, the column density
and the density of the line forming region.

The wind model described by L10 failed to produce any gas with velocities greater than $\rm{100~km~s^{-1}}$ with
a density greater than $\rm{10^{12}~cm^{-3}}$ and we come to a similar conclusion. Since L10 were trying to 
replicate the properties of the outflow observed in GRO J1655-40, which seems to have a very high density of
$\rm{5\times10^{15}cm^{-3}}$ \citep{2006Natur.441..953M} the conclusion was that the model failed in that aim.
Follow up work reduced the density estimate to around $\rm{10^{14}cm^{-3}}$ 
\citep{2008ApJ...680.1359M,2009ApJ...701..865K} but 
this is still well above the densities seen in the L10 model. This failure of a thermal wind model to replicate
the velocity and density seen in the observations suggests that
for GRO J1655-40 at least, a thermal wind seems an unlikely source of the observed absorbing gas. Nonetheless,
the predicted wind produces a column density in line with observations, and a mass loss rate in excess of the 
accretion rate.

The enhanced luminosity version of the L10 model, case Ah, does increase the velocity and perhaps
more importantly the density of the
wind. A radial velocity of $\rm{200~km~s^{-1}}$ of gas with a number density greater than $\rm{10^{12}~cm^{-3}}$ 
is still slow and less dense compared to the measurements of GRO J1655-40 mentioned above, but 
it is not unreasonable to think that this wind would produce observable features, and further work to
characterize the ionization state of this wind allowing calculation of detailed spectra would be worthwhile. The ionization 
parameter  for the fastest moving parts of the wind has a narrow range, centered on ${\log{\xi}\sim4}$, and is certainly 
similar to that inferred for the absorbing gas in  many systems.

Case
B provides the best illustration of how simple changes to the heating and cooling rates in a simulation can affect the velocity
and density of the resulting wind. We reduced line cooling by a factor of 5, and increased 
the photoionization heating rate by a factor of 4. Both of these changes can be broadly justified, by the effects of 
line optical depth in the first case, and changes to SED and gas metallicity in the second. This simple change 
made the gas thermally unstable at ${\xi}$ one order of magnitude lower. The radial location of the unstable
gas is largely unchanged, so the change in ${\xi}$ means that \emph{denser} gas is 
accelerated, producing a denser and faster wind. Although the density and velocity is only a little higher than case Ah, 
much more interesting is the huge increase in mass loss rate, now 40 times the accretion rate (even though we only
simulate a relatively small domain). This is almost 3 times the 
rate that \cite{1986ApJ...306...90S} showed would induce instabilities in the disk. Therefore, even if 
thermal winds are unable to reproduce the observed line absorption seen in XRBs,
 they may well provide a mechanism for XRB state change and so searching for an observational signature is 
 a worthwhile exercise. 
 
Another interesting result from these simulations is that there is a gas with `forbidden' 
values of ${\xi}$, i.e. from the second, hotter, unstable zone of the stability curve in cases E and F. It has been suggested
 \citep[e.g.][] {2013MNRAS.436..560C} that species which are expected to have peak abundances in gas
 with such forbidden ionization parameters would not be seen in observations. Whilst our results do not
 necessarily disprove such assumptions, they do illustrate that hydrodynamic effects (i.e. adiabatic expansion)
 make the situation more complex.
 
It is often assumed that disk winds in XRBs are equatorial, because absorption is seen preferentially in 
 sources which exhibit dips \cite{2012MNRAS.422L..11P}. We find that the wind is in fact bipolar, 
but the outflow is highly stratified with the high density region of the wind near the disk. 
 Therefore, in our simulation, absorption is only seen for equatorial sight lines even though the 
 wind flows out over a relatively wide range of angles.
 Similar results were 
 seen in L10 and \cite{2012ApJ...758...70G} who computed line profiles based on L10's as well as other disk
 wind simulations.
 This stratification is also important with respect to the observed variability of X-ray absorption in XRBs 
 \citep[][and refs therein]{2013AdSpR..52..732N}. 
This is very well illustrated
 by cases E and F, both of which produce absorbing gas in narrow angular ranges. If the illuminating spectrum 
 is variable, the angle at which particular species would be seen could change - thereby making the absorption lines 
 associated with those species vanish. The wind could remain strong in this case, but would need to be detected
 in different species.
 
 Secondly, when we compare cases A and Ah, which differ only in luminosity,
 we see that the density of the wind solution changes significantly giving rise to a very similar ionization parameter. 
This also has relevance to studies of variable sources, where increases in luminosity are sometimes called upon to explain 
increases in ${\xi}$ and hence the disappearance of some features \citep[e.g.][]{2014A&A...571A..76D}. 
Our results show that it may be  overly simplistic to assume the density remains constant in such cases, and a more
detailed investigation is required, taking into account how the wind responds to the increase in luminosity.

\section{Future work}
The simulations we have presented here use a simplified heating and cooling scheme, which permits swift
exploration of parameter space. In addition, radiative transfer through the wind is treated in the optically
thin limit. Previous detailed analysis of such hydrodynamic models \citep{sim_proga_10,2014ApJ...789...19H}
have shown that a more thorough treatment of radiative transfer including scattering can have a significant effect. 
We therefore
plan to run such simulations on the more successful models   
from this work (i.e. those that produced high velocity, dense flows). Not only will this work give more 
information regarding the validity of our modified heating/cooling rates, but it will also produce detailed ionization
data for the wind and spectra. It will also provide information regarding line emission contribution.

\section*{Acknowledgements}
 This work was supported by NASA under Astrophysics Theory Program
grants NNX11AI96G  and NNX14AK44G. The authors would like to thank the 
anonymous referee for very useful comments that have improved the paper.

\bibliographystyle{mn2e}

 \label{lastpage}

\end{document}